\documentclass[multphys,vecphys]{svmult}

\usepackage{makeidx}   
\usepackage{graphicx}  
\usepackage{multicol}  

\usepackage{cite}

\makeindex             

\begin{document}

\title{Fisher Waves and the Velocity of Front Propagation in a
Two-Species Invasion Model with Preemptive Competition}

\titlerunning{Fisher Waves in a Two-Species Invasion Model}

\author{L. O'Malley\inst{1}, B. Kozma\inst{1}, G. Korniss\inst{1},
Z. R\'acz\inst{2}
\and
T. Caraco\inst{3}}

\authorrunning{O'Malley et al.}

\institute{
Department of Physics, Applied Physics, and Astronomy, \\
Rensselaer Polytechnic Institute, 110 8th Street, Troy, NY
12180-3590, USA \\
\texttt{E-mail: omalll@rpi.edu, kozmab@rpi.edu, korniss@rpi.edu}
\and
Institute for Theoretical Physics - HAS, \\
E\"otv\"os University, P\'azm\'any s\'et\'any 1/a, 1117 Budapest, Hungary \\
\texttt{E-mail:racz@general.elte.hu}
\and
Department of Biological Sciences, \\
University at Albany, Albany NY 12222, USA \\
\texttt{E-mail: caraco@albany.edu}
}

\maketitle

\begin{abstract}
We consider an individual-based two-dimensional spatial model with
nearest-neighbor preemptive competition to study front propagation
between an invader and a resident species. In particular, we
investigate the asymptotic front velocity and compare it with
mean-field predictions.
\end{abstract}

\section{Introduction and Model}

The dynamics of propagating fronts are fundamental in the study of
the spread of advantageous alleles, species \cite{MURRAY_03}, or
opinions \cite{EBM_05}. Most notably, Fisher \cite{FISHER_37} and
Kolmogorov et al. \cite{KOLMOGOROV_37} first addressed the
velocity characteristics of a simple front by way of a
reaction-diffusion equation \cite{MURRAY_03}, which served as a
one-dimensional model for the spread of a favorable gene.
Our study of front propagation envisions introduction of an
advantageous allele or a competitively superior species through
mutation within \cite{YKC_05,OBYKAC_05} or geographic dispersal to
\cite{KC_JTB05,OAKC_SPIE} a resident population, respectively.
Introductions occur rarely, and stochastically in both space and
time.
We have shown \cite{KC_JTB05,OAKC_SPIE,YKC_05,OBYKAC_05} that the
time evolution of such systems can be well described within the
framework of homogeneous nucleation and growth. In particular, in
two dimensions, for sufficiently large systems, the typical time
scale (lifetime) until competitive exclusion of the weaker allele
or species scales as $\tau$$\sim$$(Iv^2)^{-1/3}$, where $I$ is the
stochastic nucleation rate per unit area of the successful
clusters of the better competitor and $v$ is the asymptotic radial
velocity of the corresponding circular fronts. It is, thus, clear
that the full understanding of the dependence of the lifetime on
the local rates of the systems requires that of the velocity of
the front separating the two alleles or species. Therefore, we
focus on the velocity characteristics of a two-dimensional
two-species model of invasion with preemptive competition, where
the invading species has a reproductive advantage over the
residents.

Here we study the velocity of the invading fronts of both planar
and circular shapes on $L_x$$\times$$L_y$ two-dimensional lattices
as a function of the reproduction and mortality rates of each
species, and compare them with mean-field predictions. Each
lattice site can be empty or occupied by the resident or the
invader. A lattice site represents the minimal amount of locally
available resources which can sustain an individual. Competition
for resources is preemptive, and therefore an individual site
cannot be taken by either species until its occupant's mortality
makes it available. Preemptive competition is typical for plant
species competing for common limiting resources
\cite{SHURIN_04,AMARA_rev_03,TANEY_00,YU_01,Oborny_05}. The local
occupation number at site ${\bf x}$ is $n_{i}({\bf x})=0,1$,
$i=1,2$, representing the number of resident and invader species,
respectively. Due to the excluded volume constraint, $n_{1}({\bf
x})n_{2}({\bf x})=0$, since two species cannot simultaneously
occupy the same site. A species can colonize open sites through
{\em local} clonal propagation only. An individual of either
species occupying site ${\bf x}$ may reproduce only if one or more
of its neighboring sites is empty (here we consider
nearest-neighbor interactions only).

Our unit time is one Monte Carlo step per site (MCSS) during which
$L_xL_y$ sites are chosen randomly. The occupancy (local
configuration) of a chosen site is updated based on the following
transition rates. When a site is empty, it can become occupied by
species $i$ of a neighboring site, at rate $\alpha_i\eta_i({\bf
x})$, where $\alpha_i$ is the individual-level reproduction rate
and $\eta_i({\bf x})=(1/4)\sum_{{\bf x'}\epsilon {\rm nn}({\bf
x})}n_{i}({\bf x'})$ is the density of species $i$ around site
${\bf x}$; ${\rm nn}({\bf x})$ is the set of nearest neighbors of
site ${\bf x}$. If a site is occupied by an individual, it can die
at rate $\mu$ (regardless of the species). We summarize the local
transition rules for an arbitrary site ${\bf x}$ as
\begin{equation}
0\stackrel{\alpha_1\eta_1(\bf{x})}{\longrightarrow}1, \;\;
0\stackrel{\alpha_2\eta_2(\bf{x})}{\longrightarrow}2, \;\;
1\stackrel{\mu}{\longrightarrow}0, \;\;
2\stackrel{\mu}{\longrightarrow}0, \;\;
\label{rates}
\end{equation}
where $0,1,2$ indicates whether the site is empty, or occupied by
the resident, or invader species, respectively. We study the
regime where $\mu<\alpha_1<\alpha_2$, so that competition between
the two alleles drives the dynamics  and the invading species has
an individual-level reproductive advantage over the resident.

For planar fronts, we consider an $L_x$$\times$$L_y$ lattice with
periodic boundary conditions along the $y$ direction. The
direction of propagation is along the $x$ direction by virtue of
the initial condition; we set a flat front separating the invader
species from the residents; for simplicity, to the left (right) of
the front, all sites are occupied by the invaders (residents). As
the simulation begins, a number of individuals die in a few time
steps, and in both domains, away from the front, the densities
quickly relax to their ``quasi-equilibrium" value where the clonal
propagation is balanced by mortality. The competition between the
two species, hence, takes place in the interfacial region.
Throughout the simulation, we keep track of the location of the
invading front, by defining the edge as the right-most location of
an individual of species 2, for each row of our lattice. The
average position of the front is then recorded for each time step,
from which one can extract the front velocity. We also studied the
velocity of circular fronts on an $L$$\times$$L$ lattice, with an
initial condition of a sufficiently large central cluster of the
invading species (with radius slightly larger than a critical
radius \cite{OAKC_SPIE,OBYKAC_05}), with all other sites occupied
by the resident species. We then keep track of the time-dependent
global density of the invaders, from which extracted the average
radial velocity of the growing circular cluster. Before discussing
our simulation results, we first consider those obtained from the
mean-field equations of motion.

\section{Mean-Field Equations and Propagation into an Unstable State}

\begin{figure}[t]
\centering
\includegraphics[width=.48\textwidth,height=4.7truecm]{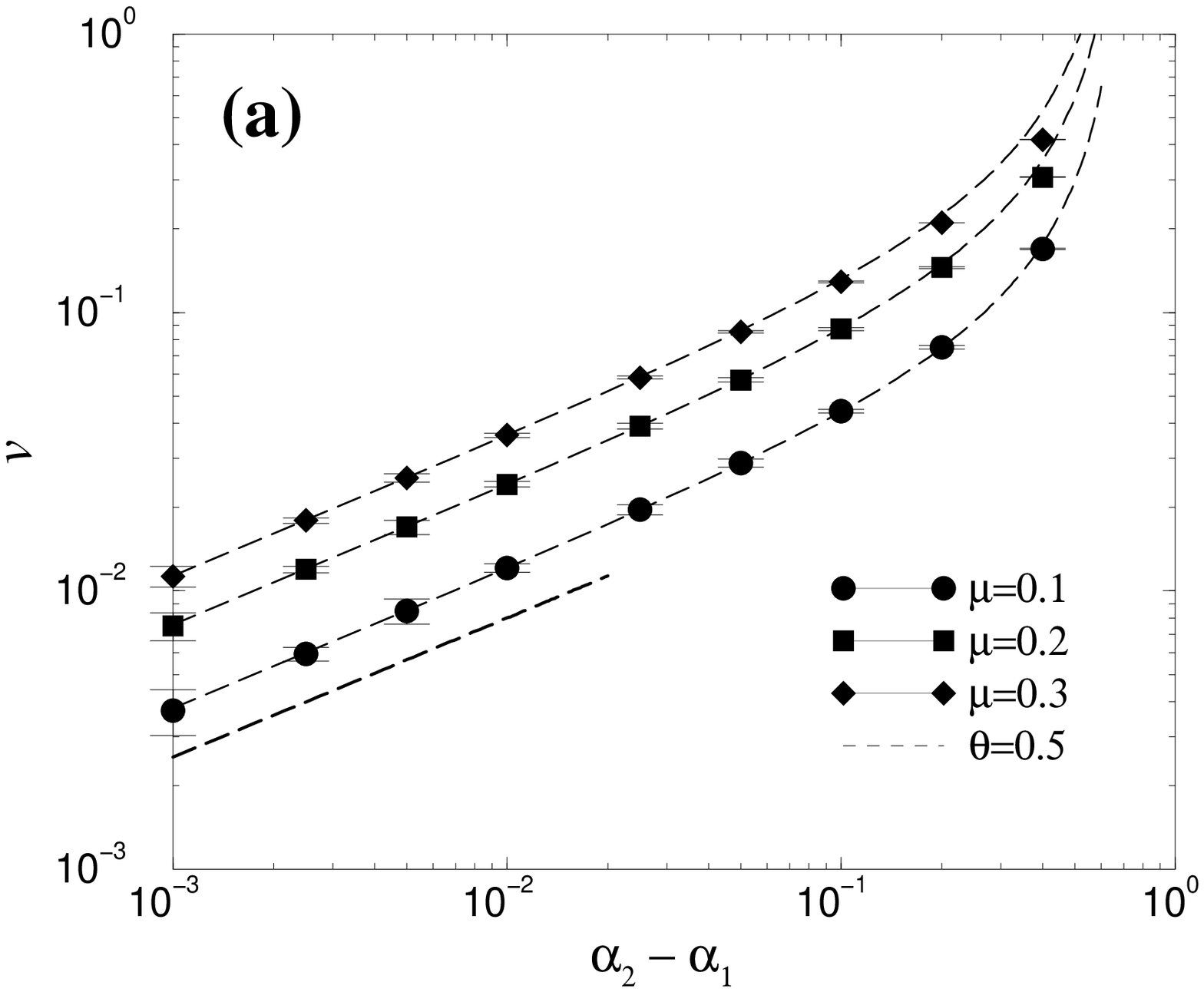}
\hspace*{0.2truecm}
\includegraphics[width=.48\textwidth,height=4.7truecm]{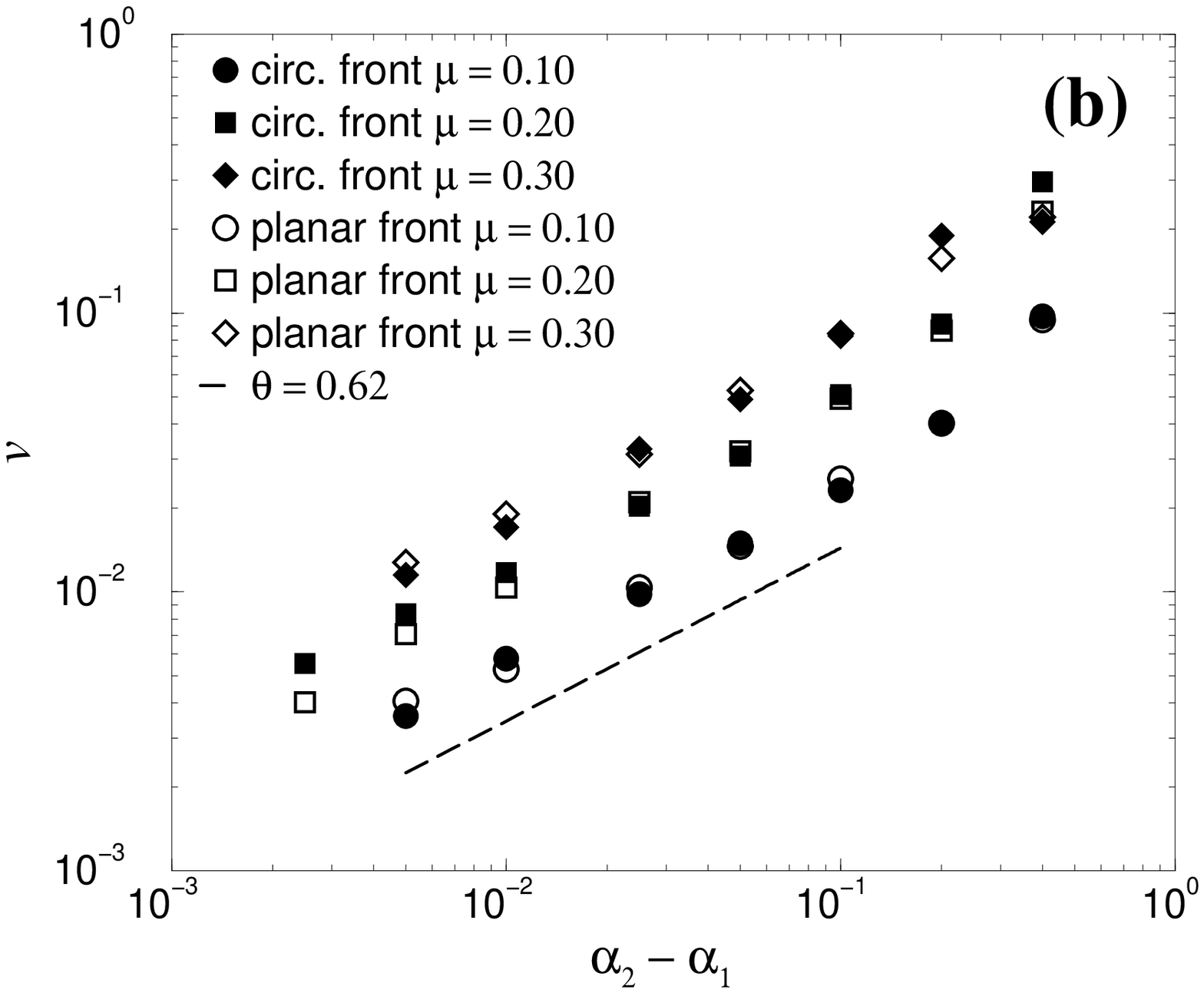}
\caption[]{ ({\bf a}) Front velocity obtained by numerical
iteration of the mean-field equations (\ref{mfe}), for fixed
$\alpha_2$=$0.70$, as a function of the difference of propagation
rates, $\alpha_2$$-$$\alpha_1$, for three different values of
$\mu$. The dashed curves are the analytic velocities of the
``marginally stable" fronts, given by Eq.~(\ref{velocity}); the
dashed straight line segment corresponds to the slope
$\theta$$=$$0.5$.
({\bf b}) Front velocity from Monte Carlo simulations for fixed
$\alpha_2$=$0.70$ as in (a), with $L_{x}=1000$, $L_{y}=100$, for
planar, and $L_{x}$$=$$L_{y}=1000$ for circular fronts, for three
values of $\mu$. The dashed straight line segment corresponds to
the effective power law with an exponent $\theta$$\approx$$0.62$
for small differences between the local reproduction rates.}
\label{fig1}
\end{figure}

From the above transition rates Eq.~(\ref{rates}) and the
underlying master equation, neglecting correlations between the
occupation numbers at different sites, for the ensemble-averaged
local densities $\rho_i({\bf x},t)$$\equiv$$\langle n_i({\bf
x},t)\rangle$ one obtains
\begin{eqnarray}
\rho_i({\bf x},t+1) - \rho_i({\bf x},t) & = & \left[1 -
\rho_1({\bf x},t) - \rho_2({\bf x},t)\right]
\frac{\alpha_i}{4}\sum_{{\bf x'}\epsilon {\rm nn}({\bf
x})}\rho_{i}({\bf x'},t) \nonumber \\
 & - & \mu \rho_i({\bf x}, t)\;,
\label{mfe}
\end{eqnarray}
$i=1,2$. For further insight we take a naive continuum limit of
Eq.~(\ref{mfe}), yielding
\begin{eqnarray}
\partial_t \rho_i({\bf x},t) & = &
\frac{\alpha_i}{4} \left[1 - \rho_1({\bf x},t) - \rho_2({\bf
x},t)\right] \nabla^2\rho_i({\bf x},t)
\nonumber\\
& + &  \alpha_i \left[1 - \rho_1({\bf x},t) - \rho_2({\bf
x},t)\right] \rho_i({\bf x},t) - \mu \rho_i({\bf x},t)\;.
\label{cmfe}
\end{eqnarray}
Spatially homogeneous solutions of these equations,
$(\rho_1^*,\rho_2^*)$, are $(0,0)$, $(1-\mu/\alpha_1,0)$, and
$(0,1-\mu/\alpha_2)$. For $\mu<\alpha_1<\alpha_2$, only the
$(0,1-\mu/\alpha_2)$ fixed point is stable. Thus, the motion of
the invading front amounts to propagation into an unstable state
\cite{FISHER_37,KOLMOGOROV_37,Aronson_78,Dee_83,Saarloos_87}, a
phenomenon that has generated a vast amount of literature
\cite{Saarloos_03} since the original papers by Fisher
\cite{FISHER_37} and Kolmogorov et al. \cite{KOLMOGOROV_37}, with
applications ranging from reaction-diffusion systems
\cite{Riordan_95,bA_98}, population dynamics \cite{MURRAY_03},
epidemics \cite{WARREN_01} or opinion formation in social systems
\cite{EBM_05}. At the level of the mean-field equations, the front
is ``pulled" by its leading edge, and for sufficiently sharp
initial profiles, the asymptotic velocity $v$ is determined by
this infinitesimally small density of invaders intruding into the
linearly unstable, resident-dominated regime. Performing standard
analysis on Eqs.~(\ref{cmfe}), one obtains the velocity of the
``marginally" stable invading fronts
\cite{MURRAY_03,Dee_83,Saarloos_87,Saarloos_03}
\begin{equation}
v^{*} = \frac{\mu}{\alpha_1} \sqrt{\alpha_2 (\alpha_2 - \alpha_1)}.
\label{velocity}
\end{equation}
Thus, for small differences in the local reproduction rates,
$v^{*}$$\sim$$(\alpha_2-\alpha_1)^{\theta}$ with $\theta=1/2$.
Equation~(\ref{velocity}) reproduces the velocity obtained by
numerically iterating the discrete-time discrete-space continuous-density mean-field
equations (\ref{mfe}) remarkable well, as shown in
Fig.~\ref{fig1}(a).

\section{Monte Carlo Results and Discussion}

We performed dynamic Monte Carlo simulations using the local rates
given by Eq.~(\ref{rates}). In the case of planar fronts, we found
that the front velocity is much smaller than that of the
mean-field approximation [Fig.~\ref{fig1}(b)]. Further, for small
differences between the local reproduction rates,
$v^{*}$$\sim$$(\alpha_2-\alpha_1)^{\theta}$ with
$\theta\approx0.62$, an exponent significantly differing from the
mean-field scaling [Fig.~\ref{fig1}]. Results from the simulations
of the propagation of circular fronts closely match those of the
planar fronts. A recent study \cite{WARREN_01} has found a similar
behavior in a discrete two-dimensional stochastic epidemic model.

The discreteness of the individuals (or equivalently, effective
density cutoffs in a continuum description)
\cite{Brunet_97,Kessler_98a,Kessler_98b} and noise
\cite{Doering_03,Doering_05} have been shown to produce velocity
characteristic drastically different from those of the mean-field
equations. More precisely, advancing fronts in stochastic
individual-based or particle models, which in the mean-field limit
converge to a pulled front behavior, are instead ``pushed"
\cite{Saarloos_03}. That is, the front velocity is determined by
the full non-linearity of the frontal region, as opposed to the
infinitesimally small leading edge \cite{Saarloos_03}. Our model
provides an example for this generic behavior.

\section*{Acknowledgments}
G.K. is grateful for discussions with Eli Ben-Naim and for the
hospitality of CNLS at the Los Alamos National Laboratory, where
some of this work was initiated. This research was supported in
part by the US NSF through Grant Nos.\ DEB-0342689 and
DMR-0426488. Z.R. has been supported in part by the Hungarian
Academy of Sciences through Grant OTKA-T043734.

%


\begin{thebibliography}{99.}
\addcontentsline{toc}{section}{References}


\bibitem{MURRAY_03}
J.D. Murray:
{\it Mathematical Biology I and II, 3rd edition} (Springer-Verlag, New York, 2002, 2003)

\bibitem{EBM_05}
E. Ben-Naim: Europhys. Lett. {\bf 69}, 671 (2005)

\bibitem{FISHER_37}
R.A. Fisher: Annals of Eugenics {\bf 7}, 355 (1937)

\bibitem{KOLMOGOROV_37}
A.N. Kolmogorov, I. Petrovsky, N. Piskounov: Moscow Univ. Bull.
Math. {\bf 1}, 1 (1937)


\bibitem{YKC_05}
J.A. Yasi, G. Korniss, T. Caraco: in {\it Computer Simulation
Studies in Condensed Matter Physics XVIII}, edited by D.P. Landau,
S.P. Lewis, and H.-B. Sch\"uttler, Springer Proceedings in Physics
(Springer-Verlag, Berlin, 2006, in press).

\bibitem{OBYKAC_05}
L. O'Malley, J. Basham, J.A. Yasi, G. Korniss, A. Allstadt, T.
Caraco: submitted to Theor. Popul. Biol. (preprint, 2005);
arXiv:q-bio/0602023

\bibitem{KC_JTB05}
G. Korniss, T. Caraco: J. Theor. Biol. {\bf 233}, 137 (2005)

\bibitem{OAKC_SPIE}
L. O'Malley, A. Allstadt, G. Korniss, T. Caraco: in {\it
Fluctuations and Noise in Biological, Biophysical, and Biomedical
Systems III}, edited by N.G. Stocks, D. Abbott, and R.P. Morse,
Proceedings of SPIE Vol. 5841 (SPIE, Bellingham, WA, 2005) pp.
117--124.


\bibitem{SHURIN_04}
J.B. Shurin, P. Amarasekare, J.M. Chase, R.D. Holt, M.F. Hoopes,
M.A. Leibold: J. Theor. Biol. {\bf 227}, 359 (2004)

\bibitem{AMARA_rev_03}
P. Amarasekare: Ecol. Lett. {\bf 6}, 1109 (2003)

\bibitem{YU_01}
D.W. Yu, H.B. Wilson: Am. Nat. {\bf 158}, 49 (2001)

\bibitem{TANEY_00}
D.E. Taneyhill: Ecol. Monogr. {\bf 70}, 495 (2000)

\bibitem{Oborny_05}
B. Oborny, G. Mesz\'ena, G. Szab\'o: Oikos {\bf 109}, 291 (2005)


\bibitem{Aronson_78}
D.G. Aronson, H. F. Weinberger: Adv. Math. {\bf 30}, 33 (1978)

\bibitem{Dee_83}
G. Dee, J. S. Langer: Phys. Rev. Lett. {\bf 50}, 383 (1983)

\bibitem{Saarloos_87}
W. van Saarloos, Phys. Rev. Lett. {\bf 58}, 2571 (1987)

\bibitem{Saarloos_03}
W. van Saarloos: Phys. Rep. {\bf 386}, 29 (2003)

\bibitem{Riordan_95}
J. Riordan, C.R. Doering, D. ben-Avraham: Phys. Rev. Lett. {\bf
75}, 565 (1995)

\bibitem{bA_98}
D. ben-Avraham: Phys. Lett. {\bf 247}, 53 (1998)

\bibitem{WARREN_01}
C.P. Warren, G. Mikus, E. Somfai, L.M. Sander: Phys. Rev. E {\bf 63}, 056103 (2001)


\bibitem{Brunet_97}
E. Brunet, B. Derrida: Phys. Rev. E {\bf 56}, 2597 (1997)

\bibitem{Kessler_98a}
D.A. Kessler, Z. Ner, L.M. Sander: Phys. Rev. E {\bf 58}, 107 (1998)

\bibitem{Kessler_98b}
D.A. Kessler, H. Levine: Nature {\bf 394}, 556 (1998)

\bibitem{Doering_03}
C.R. Doering, C. Mueller, P. Smereka: Physica A {\bf 325}, 243 (2003)

\bibitem{Doering_05}
J.G. Conlon, C.R. Doering: J. Stat. Phys. {\bf 120}, 421 (2005)



\end{thebibliography}
\end{document}